# Pit nucleation in the presence of 3D islands during heteroepitaxial growth

Mathieu Bouville[*], Joanna Mirecki Millunchick and Michael L. Falk

Department of Materials Science and Engineering

University of Michigan, Ann Arbor, MI  48109-2136

We present a model in which pit nucleation in thin films is considered to arise from a near-equilibrium nucleation process. In this model the adatom concentration plays a central role in controlling the morphological development of the surface. Although pits relieve elastic energy more efficiently than islands, pit nucleation can be prevented by a high adatom concentration. Three-dimensional islands act as adatom sinks and the lower adatom density in their vicinity promotes pit nucleation. Thermodynamic considerations predict several different growth regimes in which pits may nucleate at different stages of growth depending on the growth conditions and materials system. However direct comparisons to experimental observations require that kinetics be taken into account as well. The model predicts a wide range of possible morphologies: planar films, islands alone, islands nucleation followed by pit nucleation, and pits alone. The model shows good agreement with experimental observations in III-V systems given the uncertainties in quantifying experimental parameters such as the surface energy.

---

[*] Current address: Institute of Materials Research and Engineering (IMRE), Singapore 117602

# 1. INTRODUCTION

Morphological features such as three-dimensional islands and strain-induced ripples[1,2] arise spontaneously during the heteroepitaxial growth of III-V compound semiconductor alloys. While models exist that can explain nucleation of 3D islands or development of surface instabilities during growth,[3-6] the spontaneous formation of more complex morphologies are a matter of current investigation. Better prediction and control of such structures could improve the performance of devices and provide processing routes for the manufacture of new devices that exploit the unique physics of features near the atomic scale.

In previous work,[7,8] the authors observed the onset of pits subsequent to the nucleation of 3D islands during the heteroepitaxial growth of InGaAs on GaAs. Pits have also been observed to play an important role in morphological development in silicon-germanium on silicon,[9-11] InGaAs on InP[12] and InSb on InAs.[13] In these systems, the process of island and pit nucleation leads to surface patterning in which features on the surface are correlated to approximately 150 nm. It has been shown that pits may act as nucleation sites for islands,[11,14] resulting in island distributions and densities suitable for application such as cellular automata.[15]

These experimental observations form the basis of the theoretical analysis presented here, in which we consider the nucleation of a secondary feature, in this case a pit, on a surface upon which primary features, 3D islands, have already nucleated. We conjecture that a nucleation model is more appropriate for this material system than a linear instability model because the initial 3D island features are observed to emerge isolated on an otherwise nearly flat substrate. However it is, in part, this conjecture that we hope to evaluate by providing a detailed account of the logical consequences of such an assumption.

The central question we address here is why and how pits nucleate only at later stages of growth and near islands. There are two primary effects that could account for these inhomogeneities: a stress concentration near three-dimensional islands and a variation in adatom concentration on the surface.[16] The former effect has been studied using the finite element method,[17, 18] but no analytical form of the strain in the film exists, which makes the integration of these results difficult. The latter effect, to our knowledge, has not been addressed in detail and provides the subject of our analysis. Therefore we will, in general, assume a uniform strain across the surface. As will be discussed in section 2 one of the motivations for our focus on the role of adatom concentration is that direct experimental observations of adatoms on GaAs surfaces have revealed adatom concentrations significantly higher than was generally expected.[19] It has been theorized that this occurs due to the role of As overpressure on the surface thermodynamics.[20] Other experiments in SiGe found the adatom concentration to be almost uniform across the surface, yet small inhomogeneities in the adatom concentration were observed to lead to localization of the nucleation of islands.[21] Using a nucleation model, we show that under a variety of conditions such small inhomogeneities could account for pit nucleation, particularly in the presence of 3D islands.

## 2. THE ENERGETICS OF ISLAND AND PIT NUCLEATION

The nucleation and growth of islands and pits is driven by the ability of such features to relieve elastic energy. When an island is large enough, finite size effects can be neglected and the difference of energy between a cluster of size i and i bulk atoms can be expressed as the energy released per unit volume plus an energetic cost associated with the surface energy,

$$E_i = -\varepsilon^{isl} i + \gamma^{isl} i^{2/3}. \tag{1}$$

Here i denotes the number of atoms in the island (or the number of cation-anion pairs in the case of compound semiconductors); $\mathcal{E}^{isl}$ is an elastic energy released per atom, and $\gamma^{isl} i^{2/3}$ is the surface energy of the island minus that of a flat film. The energy of a pit obeys the same scaling,

$$E_p = - \mathcal{E}^{pit} p + \gamma^{pit} p^{2/3}, \qquad (2)$$

where p is the number of atoms (or of cation-anion pairs) which have been removed to form the pit. Even though the functional form is the same for islands and for pits, the values of $\mathcal{E}^{isl}$ and $\mathcal{E}^{pit}$ and the values of $\gamma^{isl}$ and $\gamma^{pit}$ may differ due to differences in geometry between islands and pits. In general $\mathcal{E}$ is a function of the strain, and near islands the strain is enhanced[17, 18]. However, as discussed in the introduction, the analysis presented here will focus on the effect of adatom concentrations. Therefore, the value of $\mathcal{E}$ used here will be determined from the nominal mismatch of the film.

Estimates for $\mathcal{E}$ and $\gamma$ can be obtained from the literature. Elastic calculations indicate that pits relieve strain energy more efficiently than islands.[22] *Ab initio* calculations have been used to calculate surface energies, approximately 1 eV per atom for InAs(001)[23] and GaAs(001).[24] $\mathcal{E}$ for islands and pits can be extrapolated for other materials systems by taking the mismatch and the elastic constants into account. These values lead to critical sizes, $i^*$ and $p^*$, and barrier energies, $E_i^*$ and $E_p^*$, which appear to preclude the nucleation of islands and pits, inconsistent with experimental results.

This apparent discrepancy in the barrier energy and critical size arises due to the implicit assumption that there exists an equilibrium between the island or the pit and the bulk. However when the surface is held in contact with a reservoir, e.g. arsenic overpressure during the growth of arsenides, the underlying thermodynamics of the system must reflect the exchange of adatoms with this reservoir through the chemical potential of the adatoms µ. The formation energy of an island is then expressed as

$$V_i = E_i - i\mu. \quad (3)$$

In the case of III-V semiconductors the chemical potential of a group-III adatom in the presence of an arsenic overpressure – assuming an equilibrium between island, adatoms and vapor – can be expressed as

$$\mu = E_x + kT \ln(P/P_0)/m, \quad (4)$$

where P is the pressure of the $As_m$ vapor; m is 2 if $As_2$ is used as source of arsenic and 4 for $As_4$. $P_0$ is a reference pressure and $E_x = 2.7 \pm 0.6$ eV is the formation energy of GaAs when the arsenic is initially in the vapor and Ga is an adatom.[20]

The effective critical size for island nucleation $i_{crit}$ is determined by the maximum of $V_i$,

$$i_{crit} = \frac{i^*}{(1+\mu/\mathcal{E})^3}. \quad (5)$$

If µ is larger than $\mathcal{E}$ the effective critical size can be much lower than $i^*$, which could account for the nucleation of islands on observable time scales.

In the case of pits, bonds are broken, not created, when atoms leave a pit; Eq. (3) becomes

$$V_p = E_p + p\mu. \quad (6)$$

This leads to an effective critical size of

$$p_{crit} = \frac{p^*}{(1-\mu/\mathcal{E})^3}. \quad (7)$$

Thus while the critical island size decreases when the chemical potential is taken into account the critical size for pit nucleation increases with the chemical potential. If µ is larger than $\mathcal{E}$ pits cannot nucleate as Eq. (7) has no solution. Nevertheless, pits are experimentally observed indicating that other factors come into play in the nucleation and growth of pits. In particular, this result does not take the adatom concentration into account and the related entropic effects.

In order to relate the nucleation rate of a pit to the local adatom concentration, we consider a single pit surrounded by a "gas" of adatoms with which the pit can exchange atoms. The expression for the net flux of atoms to a pit of size p is of the form:

$$\frac{dp}{dt} = \nu \left( \eta_p - \eta \right) \qquad (8)$$

where $\nu$ is an attempt frequency, $\eta$ is the number density of adatoms per site and $\eta_p$ is the equilibrium value of the adatom density per site near a pit of size p. Here we have implicitly assumed that the rate at which atoms directly impinge on a pit is small compared to the rates of exchange between the pit and the adatoms: the primary contribution to the growth or dissolution of a pit is the local adatom density. Equation (8) indicates that the growth or decay of a pit is strongly dependent on the adatom density.

The concentration of adatoms in equilibrium with a pit of size p, $\eta_p$, is determined by the chemical potential and the energetics of the pit

$$\eta_p = \eta_e \exp\left( -\frac{\Delta E_p}{kT} \right). \qquad (9)$$

$\Delta E_p$ is the difference in energy between a pit of size p and a pit of size p+1. $\eta_e$ is the equilibrium adatom concentration in the absence of islands and pits defined as

$$\eta_e = \exp\left( -\frac{\mu}{kT} \right). \qquad (10)$$

Previous reports[19] indicated that at 590ºC and an arsenic overpressure of $10^{-6}$ torr the equilibrium adatom concentration on GaAs(001) in the absence of 3D features and of deposition flux, $\eta_e$, is close to 0.1 atom per site.

Equations (8) to (10) provide a means to determine the equilibrium size of a pit as a function of experimental conditions. $\Delta E_p$, the difference in energy between a pit of size p and a pit of size p+1, can be approximated as the derivative of the energy with respect to p:

$$\Delta E_p \approx \frac{dE_p}{dp} = -\mathcal{E} + \frac{2\gamma^{pit}}{3} p^{-1/3}. \tag{11}$$

Here, and in subsequent expressions, the elastic energy relieved per atom removed from a pit is labeled $\mathcal{E}$ instead of $\mathcal{E}^{pit}$ to simplify the notation. The critical pit size is then given by

$$p_{crit} = \frac{-\gamma'^3}{\ln^3(\eta/\eta_{ceil})} \tag{12}$$

where $\eta_{ceil}$ and $\gamma'$ are defined as

$$\eta_{ceil} = \eta_e \exp\frac{\mathcal{E}}{kT} \tag{13}$$

and

$$\gamma' = \frac{2\gamma^{pit}}{3\,kT}. \tag{14}$$

When $\eta$ approaches $\eta_{ceil}$, the critical size goes to infinity. Thus pits cannot nucleate or grow at adatom concentrations above $\eta_{ceil}$. Similarly no islands can nucleate or grow below a minimal value of the adatom concentration, $\eta_{floor}$. This dependence of the nucleation of islands and pits on the adatom concentration accounts for the rarity of homogeneously-nucleated pits. As long as $\eta$ is above $\eta_{floor}$ and $\eta_{ceil}$, only islands can nucleate. Once islands form, however, they act as sinks of adatoms, lowering the adatom density and, in some cases, allowing pits to nucleate and grow.

### 3. ISLAND-INDUCED INHOMOGENEITIES IN ADATOM CONCENTRATION

Our observations of pits[7, 8] indicate that they often nucleate close to islands or even surrounded by islands. This section considers the effect of the proximity of islands on pit nucleation, particularly the case of two islands with separation much smaller than their radii. In this case the islands are treated as two infinitely long parallel absorbing boundaries a distance 2 ℓ apart. The deposition rate F acts as a source of adatoms, and steps on the surface capture adatoms at a rate proportional to the difference between the local adatom concentration η and the equilibrium adatom concentration $\eta_e$. The proportionality constant $1/\tau$ is an indication of the time it takes to incorporate adatoms into steps. Fick's second law for this one-dimensional problem is

$$\frac{\partial \eta}{\partial t} = D \frac{\partial^2 \eta}{\partial x^2} - \frac{\eta - \eta_e}{\tau} + F \tag{15}$$

where x is the position between the islands and D is the diffusivity. In steady state, the adatom concentration at position x, η(x), is given by

$$\frac{\eta(x) - \eta_\infty}{\eta_{edge} - \eta_\infty} = \frac{\cosh x/L}{\cosh \ell/L} \tag{16}$$

where $L = \sqrt{D\tau}$ is a diffusion length, $\eta_{edge}$ is the adatom concentration at the island edge and $\eta_\infty = \eta_e + F\tau$ is the adatom concentration in the absence of islands. All distances are expressed in terms of distance between two surface sites, areas are in number of surface sites. Figure 1 shows how η varies between two islands separated by 2 ℓ as a function of position as given by Eq. (16). The adatom concentration is minimum at the edge of the island where $\eta = \eta_{edge}$ and reaches its maximum value midway between the islands, $\eta_{mid}$, which is given by

$$\frac{\eta_{mid} - \eta_\infty}{\eta_{edge} - \eta_\infty} = \text{sech} \frac{\ell}{L} \tag{17}$$

If $\eta_{ceil} < \eta_{edge}$, the adatom concentration is greater than $\eta_{ceil}$ everywhere on the surface and pitting is precluded. When $\eta_{edge} < \eta_{ceil} < \eta_{mid}$, pitting is localized near the islands. When $\eta_{ceil} > \eta_{mid}$, the adatom concentration is lower than $\eta_{ceil}$ everywhere between the islands and pit nucleation is delocalized. Finally when $\eta_{ceil} > \eta_\infty$, the adatom concentration is lower than $\eta_{ceil}$ even in the absence of islands and pits can nucleate.

Whether or not pits may nucleate thus depends on the value of $\eta_{edge}$, which can be derived by considering the conservation of mass on the surface. In steady state, this gives

$$\eta_{edge} \approx \eta_\infty - \frac{\eta_\infty - \eta_e \exp\left(\frac{\Delta E_i}{kT}\right)}{1 + \frac{1}{\Lambda}\tanh\frac{\ell}{L}} \tag{18}$$

where $\Lambda = 4 L \exp(-E_{local}/kT)$ and $E_{local}$ is an energy barrier to detachment from the island edge.

## 4. THE THERMODYNAMICS OF PIT NUCLEATION IN THE PRESENCE OF ISLANDS

The results detailed in the previous two sections allow us to present a detailed picture of when pitting is thermodynamically favored subsequent to three-dimensional islanding for a particular material system under given growth conditions. When pit nucleation is thermodynamically allowed pits may only be thermodynamically possible in the vicinity of islands or they may arise anywhere between the islands. We set aside the question of the kinetics of pit nucleation until section 5.

We begin by determining whether pitting is possible for a given materials system under a given set of growth conditions and at a stage of growth characterized by a particular island-island separation. This is accomplished by comparing the adatom concentration above which pits cannot nucleate, $\eta_{ceil}$, with the lowest adatom concentration on the surface that occurs next to the islands, $\eta_{edge}$. When $\eta_{edge}$ is greater than $\eta_{ceil}$ pitting is precluded. From Eq. (13) and (18) this condition is equivalent to

$$\tanh\frac{\ell}{L} > \Lambda \frac{\eta_{ceil} - \eta_e \exp\left(\frac{\Delta E_i}{kT}\right)}{\eta_\infty - \eta_{ceil}}. \tag{19}$$

Next we distinguish between instances in which pits can form at any arbitrary location between islands or only adjacent to the islands. When $\eta_{ceil}$ is greater than even the highest adatom concentration on the surface pitting is delocalized. Since the maximum adatom concentration is always smaller than $\eta_\infty$, pitting can occur irrespective of the presence of islands when $\eta_\infty$ is below $\eta_{ceil}$. From Eq. (13) and our definition of $\eta_\infty$ this condition is equivalent to $F\tau/\eta_e < e^{\varepsilon/kT}-1$. Note that $F\tau/\eta_e$ is the supersaturation, i.e. the relative increase of the adatom concentration due to deposition. When $\eta_\infty$ is higher than $\eta_{ceil}$, islands in proximity to each other may still decrease the maximum adatom concentration in the region between them below $\eta_{ceil}$. Using Eq. (17) and (18) the condition for delocalized pitting is then equivalent to

$$\cosh\frac{\ell}{L} + \frac{1}{\Lambda}\sinh\frac{\ell}{L} < \frac{\eta_\infty - \eta_e \exp\left(\frac{\Delta E_i}{kT}\right)}{\eta_\infty - \eta_{ceil}} \tag{20}$$

Thus when $\eta_{ceil} < \eta_\infty$, the transition between delocalized and localized pitting depends on the distance between islands, $\ell$.

Figure 2 illustrates the transition from absence of pitting to localized and then delocalized pitting as a function of $\eta_{ceil}/\eta_\infty$ and of $\ell/L$. $\eta_{ceil}$ depends on the growth rate and $\eta_\infty$ on the mismatch and temperature. Therefore the x-axis of Fig. 2 depends on the material system under certain growth conditions which we define as an "experimental regime". The y-axis depends on the distance between two islands, i.e. the stage of growth. Assuming that islands have already nucleated, $\ell$ decreases as they approach each other due to further growth. In Fig. 2 the growth process subsequent to 3D islanding can be conceived as a downward-pointing vertical arrow indicative of the time evolution of the system due to the decrease of ℓ. Therefore, from Fig. 2 we can infer the morphological evolution for a given experimental regime. If $\eta_{ceil}$ is always greater than $\eta_\infty$ pitting is delocalized. We will refer to this experimental regime as *delocalized* pitting, or "ID" for *i*slands followed by *d*elocalized pits. If $\eta_{ceil}$ is less than $\eta_\infty$ but greater than a second transition value, $\eta^0_{ceil}$ pitting is initially possible only near islands until the islands reach a critical separation. This is the *adjacent* pitting experimental regime which we will denote "IA". If $\eta_{ceil}$ is lower than $\eta^0_{ceil}$ pitting is precluded until islands are within a critical separation distance. Thus the systems with the lowest values of $\eta_{ceil}/\eta_\infty$ are designated as exhibiting pits *between* islands and are denoted "IB".

Figure 3 shows these three experimental regimes as a function of $\mathcal{E}/kT$, the elastic energy as compared to the thermal energy, and $F\tau/\eta_e$, the supersaturation. The transition between experimental regimes IB and IA corresponds to the transition between pitting in between islands and pitting adjacent to islands. If $\Lambda$ is small, this transition is close to the IA-ID transition, the region of phase space associated with experimental regimes of type IA is small. If $\Lambda$ is large on the other hand, $F\tau/\eta_e$ must be much larger than $\mathcal{E}/kT$; limiting the phase space of experimental regimes of type IB. Note that Fig. 3 assumes that islands have already nucleated. This criterion will be relaxed when kinetic effects are taken into account to determine if islands nucleate prior to pits, if at all.

### 5. THE KINETICS OF ISLAND AND PIT NUCLEATION

The previous section detailed the thermodynamics of pit nucleation. However, the epitaxial growth process occurs on experimentally-determined time scales. Pits will not be experimentally observed unless they nucleate on time scales that are comparable to or faster than this time. In order to predict the experimental observation of pits or lack thereof, it is necessary to incorporate the effect of kinetics into our theoretical analysis.

The rate at which pits nucleate is proportional to the rate at which pits of critical size are generated. The statistics of pit populations on the surface — in particular the number of pits of critical size — can be obtained using the formalism introduced by Walton.[25] Assuming detailed balance, we can derive an expression for the number of pits of size p

$$N_p = N_1 \, e^{-\frac{E_p - E_1}{kT}} \left(\frac{\eta_e}{\eta}\right)^{p-1} \tag{21}$$

where $N_1$ is the number of pits of size 1 and $E_1$ is the energy of a pit of size 1. The nucleation rate is proportional to the number of pits of critical size. Because $N_1 = \exp\left(-\frac{E_1 + \mu}{kT}\right)$, the rate at which pits nucleate is

$$R = R_0 \, \eta \exp\left(-\frac{\gamma'^3}{2 \ln^2 \eta/\eta_{ceil}}\right) \tag{22}$$

where $\gamma' = \frac{2\gamma^{pit}}{3 \, kT}$ and $R_0$ is a rate constant related to the attempt frequency. The nucleation rate is high at low $\eta/\eta_{ceil}$ and decreases rapidly with increasing $\eta/\eta_{ceil}$. The nucleation rate decreases when the surface energy increases and, for large values of $\gamma'$, only at low adatom concentrations can pits nucleate at a non-negligible rate. Moreover the maximum nucleation rate is lower at larger $\gamma'$. When $\eta/\eta_{ceil}$ is less than $e^{-\gamma'}$, $p_{crit}$ would be less than 1 and the model breaks down.

In the previous section we showed that when islands are arbitrarily far apart experimental regimes of type ID show pitting regardless of the presence of islands whereas those of type IA can nucleate pits only adjacent to the islands. An experimental regime that is thermodynamically of type ID would appear to be kinetically of type IA if the probability for a pit to nucleate far from an island is small compared to the probability that it nucleate close to an island. Therefore the ratio of the nucleation rate close to the island to the nucleation rate far from the island will be used to kinetically discriminate experimental regimes of types IA and ID. The pit nucleation rate given by Eq. (24) is highest at the islands, where the adatom concentration is the lowest,

$$R_{edge} = R_0 \, \eta_{edge} \, \exp\left(-\frac{\gamma'^3}{2\ln^2 \eta_{edge}/\eta_{ceil}}\right) \tag{23}$$

and it is lowest at the mid-point between the islands, where $\eta$ is the highest,

$$R_{mid} = R_0 \, \eta_{mid} \, \exp\left(-\frac{\gamma'^3}{2\ln^2 \eta_{mid}/\eta_{ceil}}\right). \tag{24}$$

If the ratio of these two rates $R_{edge}/R_{mid}$ is high, pits nucleate primarily adjacent to the islands as in IA. If this ratio is close to one, pits can nucleate everywhere at almost the same rate and the experimental regime is kinetically of type ID. We therefore define the *kinetic* IA-ID transition as $R_{edge}/R_{mid}$ much greater than one, which will be 100 for our purposes.

While some experiments showing pits were described in the previous section, most systems do not exhibit any pitting. This implies that there is an important experimental regime for which pitting is precluded for any value of $\ell$. In that case, the maximum nucleation rate on the surface which occurs next to the islands must be negligible regardless of island-island separation. We will denote the experimental regimes where pits are never observed to nucleate, i.e. where there are only islands, as being of type "I0". The adatom density at the island, $\eta_{edge}$, is at its lowest when $\ell = 0$. The condition for the I0-IB transition can be found by setting $\eta_{edge}$ to $\eta_e \exp\left(\frac{\Delta E_i}{kT}\right)$ in Eq. (23).

In Eq. (23), $R_{edge}$ gives the nucleation rate in number of pits per site per second. To compare our predictions to experiments, it is more convenient to express this condition in number of pits per sample. The cut-off value is again arbitrary; we will consider experimental regimes to be kinetically of type IB, in which pits nucleate only between islands, when fewer than 1 pit nucleates per minute per 100 μm$^2$; this gives

$$R_{edge}\,(\ell\rightarrow\infty)/R_0 < 10^{-23} \tag{25}$$

as a criterion. Experimental regimes are kinetically of type IA when pits can nucleate only close to islands when the latter are far apart. This corresponds to the condition that

$$R_{edge}\,(\ell\rightarrow\infty)/R_0 > 10^{-23}, \tag{26a}$$

$$R_{edge}/R_{mid} > 10^2. \tag{26b}$$

Experimental regimes are kinetically of type ID when pits can nucleate everywhere even if islands are far apart. This corresponds to the condition that

$$R_{edge}/R_{mid} < 10^2. \tag{27}$$

Experimental regimes are kinetically of type I0 when pits cannot nucleate even when islands are close together; it is defined as

$$R_{edge}\,(\ell = 0)/R_0 < 10^{-23}. \tag{28}$$

Figure 4 shows the kinetic phase diagram obtained from these conditions. The dashed lines represent the thermodynamics results for the IB-IA and IA-ID transitions from Fig. 3 for comparison. When the surface energy γ' is small, the kinetic phase diagram is very close to that from thermodynamics. But when γ' is larger, as in Fig. 4, the boundaries shift to higher values of $\mathcal{E}/kT$. This is because when the surface energy is larger, the nucleation process is slower. Experimental regimes of type ID which dominated in Fig. 3 may not exist at all on the observable time scale when γ' is large and the discrepancy between thermodynamics and kinetics is most significant. Regimes of type I0 on the other hand, which have no analog when kinetics are not considered, account for most low-misfit material systems.

The four experimental regimes we have defined so far assume that islands have already nucleated. To account for the possible absence of island nucleation and the fact that pit may nucleate before islands, we add two more experimental regimes: "P" when pits nucleate before islands and "0" when both islands and pits are kinetically prevented. The nucleation rate of islands is similar to that of pits, Eq. (22),

$$R_{isl} = R_0 \frac{\eta_e^2}{\eta_\infty} \exp\left(-\frac{\gamma'^3}{2\ln^2 \eta_{floor}/\eta_\infty}\right) \qquad (29)$$

If $R_{isl}/R_0 > 10^{-23}$, islands can nucleate on a flat film. If $R_{pit}/R_0 > 10^{-23}$, pits can nucleate on a flat film. If both rates are small, then neither islands nor pits nucleate and the film remains planar, i.e. regime "0". If both can nucleate, it is necessary to determine which one nucleates at a higher rate to discriminate between "I" and "P" regimes. To this end, we compare their nucleation rates on a flat film at $\eta = \eta_\infty$.

## 6. DISCUSSION

Experimental observations show a wide range of morphologies: planar films, islands alone, islands nucleation followed by pit nucleation, pits alone. Materials systems and experimental procedures also are very diverse; the mismatch, surface energy, temperature can all vary. In this section we discuss the effects of the changes of such parameters on the surface morphology.

Figure 5 shows the experimental regimes as a function of the surface and strain energies. The two diagrams are for two different values of the supersaturation, $F\tau/\eta_e$. At low mismatch and high surface energy, neither islands nor pits can nucleate and the film remains planar (regime 0). At the highest strain energies, pits can nucleate before islands (regime P). If both surface and strain energy are low, islands nucleate (regime I). There are four sub-regimes in regime I. At very low surface and strain energies only islands nucleate (regime I0). For higher strain energies, pits can nucleate but are limited to nucleation between islands (regime IB). For even higher strain energies, they nucleate adjacent to islands (regime IA) or pit nucleation is delocalized between the islands (regime ID).

These six experimental regimes exist at all supersaturations, but which regime dominates depends on the value of the supersaturation. Equation (12) predicts that low adatom concentrations promote pit formation. As a result, at lower supersaturations, Fig. 5(a) for instance, there are essentially two cases: either pits nucleate before islands or neither islands nor pits nucleate. At high supersaturations, Fig. 5(b), the islanding regions dominate regions where pits nucleate first. We have already shown that a high adatom concentration promotes islands formation and prevents the formation of pits; Fig. 5 indeed shows that the islanding regions expand at high supersaturations, while pitting regions shrink.

In addition to studying pit nucleation for different strain and surface energies, we have examined the effect of parameters such as deposition rate and temperature for a single materials system. Figure 6 shows the predicted experimental regimes as a function of temperature and deposition flux for a film where $\mathcal{E}/kT = 0.13$ and $\gamma' = 1.10$. The model predicts several different possible morphologies for a given set of growth conditions. For the lowest growth rates and temperatures, the film is predicted to remain planar, with neither islands nor pits able to nucleate on the surface. With increasing growth temperatures, pit nucleation alone is expected at low growth rates. As the growth rate increases, islands are expected to nucleate. However, the model invariably predicts that pitting occurs subsequent to islanding. As we will discuss in the next section, this is consistent with experimental observations in III-V systems.

# 7. COMPARISON TO EXPERIMENTS

The predictions of the model will be compared to experimental results for two materials systems. At intermediate mismatch, the model predicts a small region of phase space where pits can nucleate adjacent to islands, as shown in Fig. 5(a). Figure 7(a) shows a 20 monolayers-thick $In_{0.27}Ga_{0.73}As$/GaAs film grown at T = 500ºC and F = 2.2 Å/s the surface of which is covered with islands. Next to the islands pits are also observed. These pits form over a large range of growth conditions, but only after a significant number of islands have nucleated.[7, 8] For these growth conditions, $\mathcal{E}/kT \approx 0.13$, and the geometry of the pits indicates $\gamma \approx \gamma_{(001)}/6$. Using the value for $\gamma_{(001)}$ calculated by *ab initio* methods,[23, 24] $\gamma' \approx 1.9$. Assuming that $\eta_e = 0.1$ and $\Lambda = 1$, this film is expected to reside in the type 0 region, where neither islands nor pits can nucleate and grow, very near the boundary of type IB, where pits may nucleate between islands, as shown by the diamond in Fig. 5(b). This small discrepancy may come from the uncertainty in the elastic and surface energies. It may also arise due to the fact that the stress inhomogeneities induced by the islands lead to a local increase in the strain at the island edge. In fact, finite element calculations show that the strain close to an island[17, 18] can be twice as high as the nominal mismatch. This strain concentration at the island edge would increase $\mathcal{E}$, pushing the system into regime IB.

At higher mismatch the model predicts that pit nucleation may occur more readily, which has been observed experimentally. Figure 7(b) shows a 7 monolayers-thick InSb/InAs film grown at T = 400ºC and F ≈ 1.3 Å/s the surface of which is covered with large rectangular islands, with small pits visible between them[13]. For these growth conditions, $\mathcal{E}/kT \approx 1.35$. Surface energies are not known for InSb but we estimate that the value of $\gamma'$ for InSb/InAs is around 3.5 based upon the lower melting temperature of antimonides compared to arsenides. Thus, we predict that this film should reside in regime IA close to the IA-ID border, as denoted by the open square in Fig.

5(b), while experimentally the morphology is observed to be of type ID. This discrepancy is small considering the uncertainty on some of the parameters. Furthermore, the observed trend is correct, pitting is more favorable in InSb/InAs than in InGaAs/GaAs.

The model predicts that material systems with a high mismatch are more likely to form pits as the driving force (elastic energy relaxation) is higher. However, a high mismatch also implies a very low critical thickness. Pitting may thus be prevented by the lack of material to support a pit.[10] As the growth mode in these films is Stransky-Krastanov, denuding the substrate is not energetically favorable. For this reason, a system such as InAs/GaAs would not show any pitting in spite of its high misfit, because the critical thickness is on the order of 1 or 2 monolayers. InSb/InAs can support pit nucleation despite its high mismatch because the film and substrate differ in their group-V species. When the InAs substrate is exposed as a pit grows, the volatile arsenic atoms have a high probability of desorbing. Since the overpressure consists of Sb vapor, the layer is converted to InSb. This results in an effective wetting layer that is infinitely thick, thus allowing for unhampered pit nucleation and growth[13].

In addition to studying pit nucleation in different materials systems, we can also choose one materials system and study the effect of parameters such as deposition rate, temperature and arsenic overpressure. Figure 8 shows the expected experimental regimes as a function of growth rate and deposition temperature for $\gamma' = 1.9$ and $\mathcal{E}/kT = 0.13$, which are close to the nominal values for $In_{0.27}Ga_{0.73}As/GaAs$. For this set of parameters, planar films are expected at low growth rates over a typical range of growth temperatures. At higher growth rates the nucleation of pits between islands is observed at lower temperatures (IB), while pit nucleation adjacent to islands is seen at higher temperatures (IA). These predictions are consistent with experimental results for $In_{0.27}Ga_{0.73}As/GaAs$[7,8]. Figures 8(b) and (c) show a pair of atomic force micrographs of $In_{0.27}Ga_{0.73}As/GaAs$. The sample in Fig. 8(b) is 15 ML thick and was deposited at T = 505°C, F = 5 Å/s, and an $As_4$ overpressure of $12 \times 10^{-6}$ torr. The sample in Fig. 8(c) is 21 ML thick and was grown at T = 505°C, F = 0.7Å/s, and an $As_4$ overpressure of $16 \times 10^{-6}$ torr. For the high growth rate sample, pits are observed to nucleate between the islands. When the growth rate is decreased at the same temperature, pits are observed to nucleate adjacent to islands, consistent with the predictions of Fig. 8(a).

Experimentally, it has also been reported that island and pit nucleation depends on the arsenic overpressure such that at high arsenic overpressure, island and pit nucleation is delayed[7-8]. The arsenic overpressure is known to have an effect on the chemical potential $\mu$, the surface energy and the diffusivity, however, the dependence of the diffusivity on As overpressure is not well established. It is therefore not possible to compare our model to experiments in terms of the effect of arsenic overpressure as changes in diffusivity cannot be taken into account with any precision. The fact that high arsenic overpressure delays island and pit nucleation suggests that changes in the diffusivity are the primary effect of arsenic overpressure.

This model agrees well with the observations in compound semiconductors, but pits have also been observed in SiGe systems. Jesson *et al.* for example observe cooperative nucleation of islands and pits in $Si_{0.5}Ge_{0.5}$ grown on Si(001) substrates at low temperature and annealed for 5 min at a T = 590°C[9]. Our model predicts that for these conditions the film should remain planar, in apparent contradiction with the experimental results. However, in our theoretical treatment the assumption was made that pits and islands nucleate independently. Jesson suggested a cooperative nucleation mechanism, i.e. a simultaneous nucleation of an island-pit pair. Such a mechanism is not taken into account in our model. Gray *et al.* show that for $Si_{1-x}Ge_x$ grown on Si(001) at T = 550°C, F = 1Å/s and 25 % < x < 50 %, shallow pits are observed to nucleate prior to the nucleation of islands.[11] In contrast, our model predicts that islands should nucleate prior to pit nucleation assuming a symmetry in the aspect ratio of these features. The observations of Gray *et al.* is consistent with a morphology that may arise as a result of a localized surface instability[26] as opposed to a nucleation event. In neither of these systems does our model predict the observed morphologies, suggesting that the mechanisms which dominate in SiGe systems are different from those in compound semiconductors.

Although the predictions of the model agree well with experimental observations, an important *caveat* should be noted. There exists significant uncertainty in the values of the materials parameters due to the lack of experimental data. The strain energy relieved by the pits, $\mathcal{E}$, necessarily depends on the pit geometry, and the surface energy is not well known for all of the materials systems. In our model $\gamma\, p^{2/3}$ is the surface energy of the pit minus that of a flat film. In the case of growth on (001) surfaces

$$\gamma\, p^{2/3} = \gamma_p A_p - \gamma_{(001)} A_{(001)} \qquad (30)$$

where $A_p$ is the surface area of the pit and $A_{(001)}$ is its basal area in the (001) plane. $\gamma_p$ and $\gamma_{(001)}$ are the surface energies per unit area on the surface of the pit and (001) respectively. As surface energies for planes other than (001) are generally not known, we assume that $\gamma_p \approx \gamma_{(001)}$ in order to provide an estimate of $\gamma$. For the purposes of our analysis, we use the empirically observed pit shapes to determine these factors.

We also remind the reader that a number of important simplifying assumptions were made in constructing the theory. Foremost amongst these is that the stress is taken to be homogeneous, which is obviously not rigorously true near surface features. Stress inhomogeneities may very well play an important role in controlling the nucleation of islands or pits. However, we wish to emphasize that even without accounting for stress inhomogeneities we obtain growth regimes and trends in agreement with experiment. In addition we have restricted our investigation to a one-dimensional treatment of diffusion, and we neglect anisotropy in the geometry and the elastic and diffusive coefficients. Although including these would result in quantitatively different predictions, the simplified model we have constructed accounts for inhomogeneity of the adatom concentration and predicts the conditions under which pit nucleation is favored close to islands. As far as we are aware this is a feature typically neglected in other models. Fluctuations in the adatom concentration about the steady-state value are also neglected. This is, perhaps, the least understood aspect of the statistical physics of surfaces and clearly deserves further study, but is beyond the scope of current investigation.

The model presented here assumes a nucleation and growth mechanism. This is not meant to imply that pitting cannot arise due to other mechanisms. Jesson *et al.* for instance observed a cooperative nucleation of islands and pits[9], while our model considers only for the sequential nucleation of the features. Similarly the model obviously would not apply to cases where pits arise due to a linear instability induced by the growth conditions.[27]

## 8. CONCLUSION

We have studied the nucleation of islands and pits during heteroepitaxial growth of semiconductors. In this model pit nucleation arises from a near-equilibrium nucleation process where the adatom concentration plays a dominant role. Elastic calculations show that pits can relieve elastic energy more efficiently than islands. However their nucleation is sensitive to the adatom concentration and can be altogether prevented by a high adatom concentration. Also the inhomogeneity of the adatom concentration due to diffusion favors pit nucleation close to the islands where the adatom concentration is lower. We found that while energetic arguments indicate that pits should dominate, they are typically kinetically prevented. Taking kinetics into account, we identified six experimental regimes depending on the growth rate and the elastic energy due to the misfit: pits can nucleate far from islands, adjacent to isolated islands, in between islands, or in the absence of islands. The film can also remain planar or islands alone can nucleate. There is reasonable agreement of the theory with experiments in III-V systems given the uncertainties in quantifying experimental parameters.

## ACKNOWLEDGEMENTS

The authors wish to thank Alexandru Riposan for helpful discussions of his experimental data. This work was supported by NSF under grant DMR0092602.

**Figures captions**

FIG. 1. The variation of the adatom concentration η between two islands separated by a distance 2ℓ, as a function of position, as given by equation (16).

FIG. 2. Transitions between no pitting, localized pitting and delocalized pitting as a function of both the ratio of the critical adatom concentration for pit formation to the adatom concentration on a nominally flat film, $\eta_{ceil}/\eta_\infty$ and of the ratio of the island separation to the diffusion length, $\ell/L$. The transition between no pitting and localized pitting is obtained from Eq. (21) and the localized-delocalized transition is from Eq. (22). Drawn assuming $\Lambda = 1$. The dashed lines are asymptotes.

FIG. 3. Equilibrium phase diagram showing the domains of the experimental regimes where pits nucleate adjacent to islands (IA), pits nucleate between islands (IB) and pits are delocalized (ID) as a function of the ratio of the elastic energy to the thermal energy, $\mathcal{E}^{pit}/kT$, and of the supersaturation induced by the beam, $F\tau/\eta_e$. The IB-IA boundary is made of the roots of $\eta_{ceil} = \eta^0_{ceil}$ and the IB-ID boundary corresponds to $\eta_{ceil} = \eta_\infty$. Plotted with $\Lambda = 1$.

FIG. 4. Kinetic phase diagram showing the domains of experimental regimes where only islands nucleate (I0) obtained from Eq. (28), pits nucleate adjacent to islands (IA) from Eq. (26), pits nucleate between islands (IB) from Eq. (25) and pits are delocalized (ID) from Eq. (27) as a function of the ratio of the elastic energy to the thermal energy, $\mathcal{E}^{pit}/kT$, and of the supersaturation induced by the beam $F\tau/\eta_e$. The dashed lines correspond to the thermodynamic results shown in Fig. 2. In this graph, we have assumed $\Lambda = 1$ and $\gamma' = 2$.

FIG. 5. Domains of the various experimental regimes as a function of strain and surface energies calculated from Eqs. (25)-(28) as in Fig. 4, drawn assuming $\eta_e = 0.1$ and $\Lambda = 1$, for two different values of the supersaturation (a) $F\tau/\eta_e = 0.02$, (b) $F\tau/\eta_e = 0.07$. Two experimental systems are denoted for comparison, $In_{0.27}Ga_{0.73}As/GaAs$[7-10] (♦) and $InSb/InAs$[15] (□). Note: the geometry of pits is accounted for in the evaluation of the surface energy.

FIG. 6. Domains of the various experimental regimes calculated from Eqs. (25)-(28) as in Fig. 4, as a function of temperature and deposition rate (in arbitrary units), assuming that at 500°C, $\mathcal{E}/kT = 0.13$, $\Lambda = 1$ and $\gamma' = 1.10$.

FIG. 7. AFM images of (a) a 22 ML thick $In_{.27}Ga_{.73}As$ film grown on GaAs at 500°C and (b) a 7 ML thick InSb/InAs film grown at 400°C.

FIG. 8. (a) Domains of the various experimental regimes as a function of temperature and growth rate calculated from Eqs. (25)-(28) as in Fig. 4, in arbitrary units for $\gamma' = 1.9$, $\mathcal{E}/kT = 0.13$, $\eta_e = 0.1$ and $\Lambda = 1$. AFM micrographs of $In_{0.27}Ga_{0.73}As/GaAs$ grown at T = 505°C and As overpressure = $12.10^{-6}$ torr at (b) F = 1.75 ML/s, h = 15 ML and (c) F = 0.25 ML/s, h = 21 ML[8]. The arrow in (b) points to a pit that has nucleated between a cluster of islands.

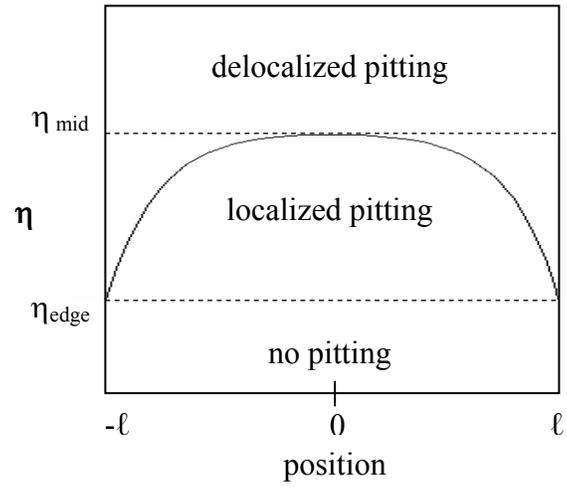

Figure 1

Bouville *et al.*

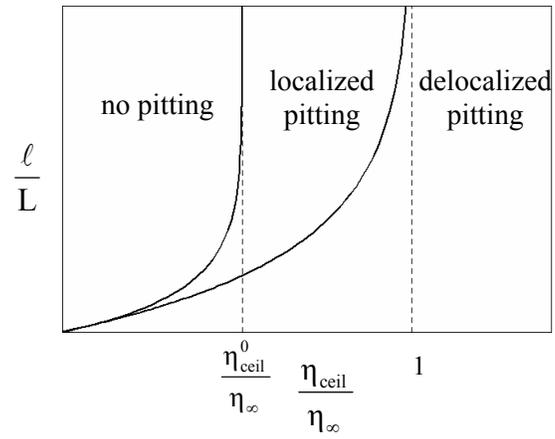

*Figure 2*

Bouville *et al.*

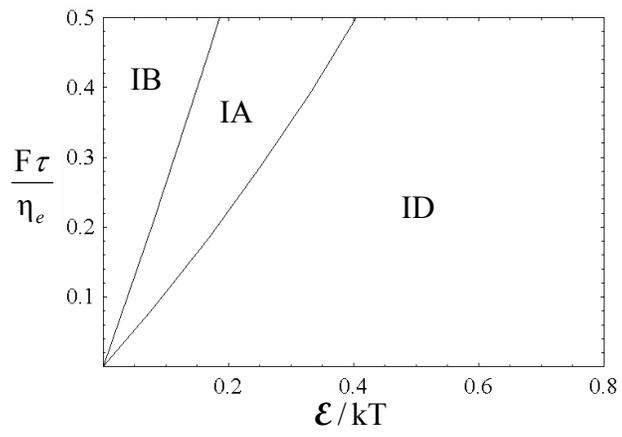

*Figure 3*

Bouville *et al.*

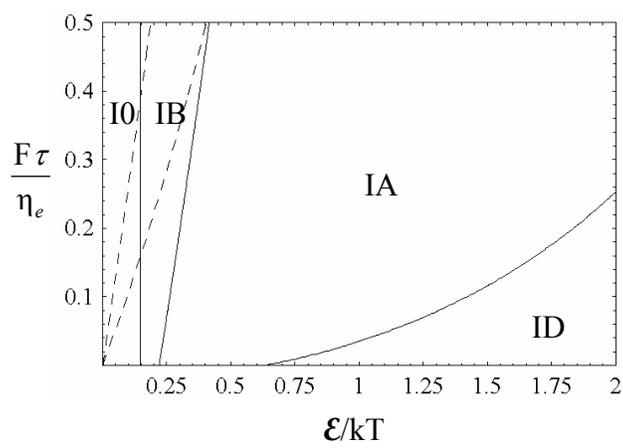

*Figure 4*

Bouville *et al.*

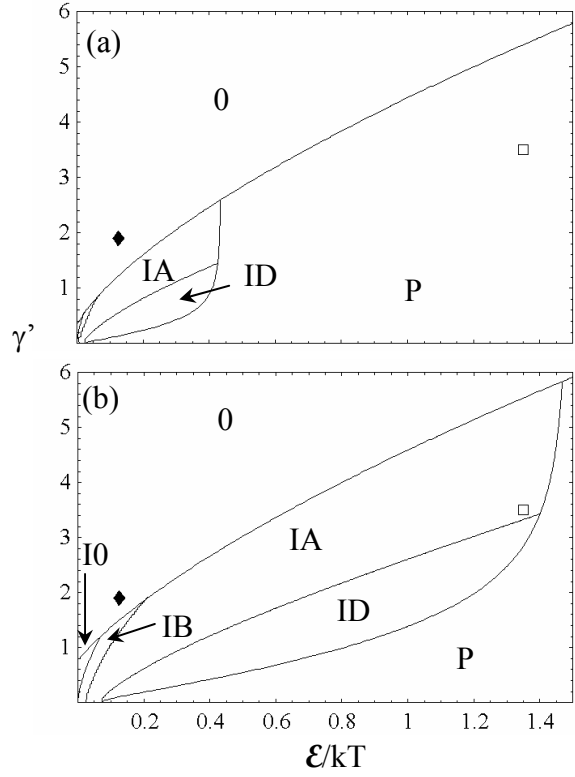

Figure 5

Bouville *et al.*

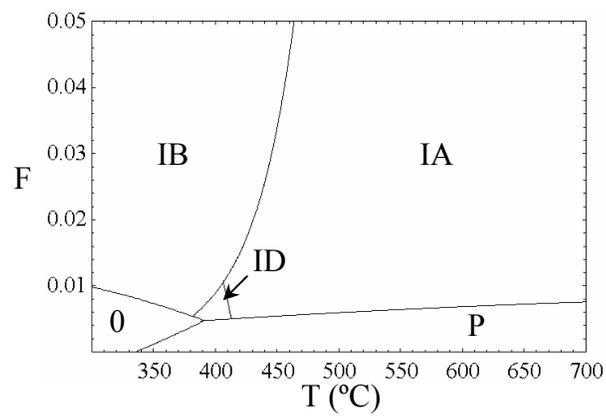

Figure 6
Bouville *et al.*

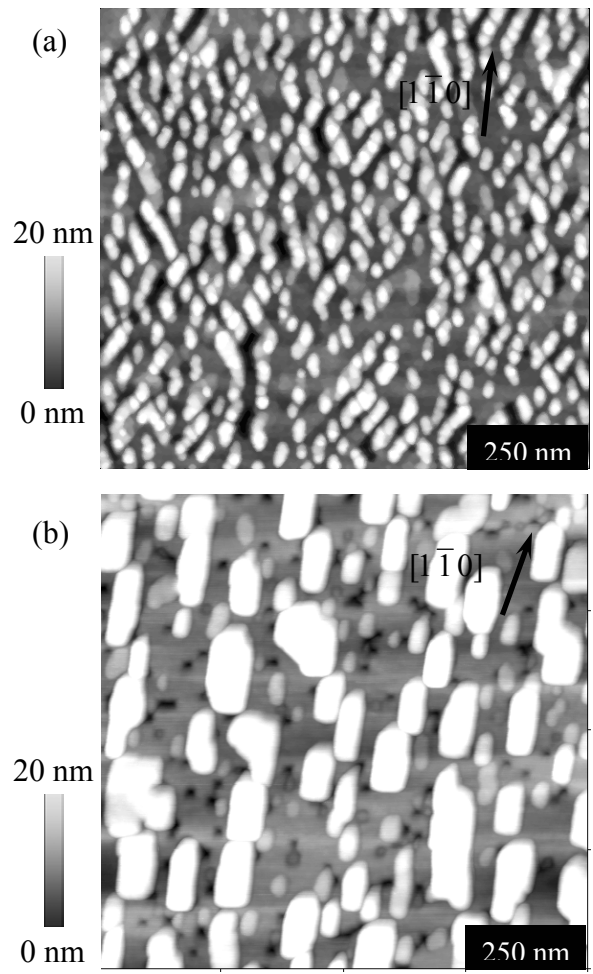

Figure 7

Bouville *et al.*

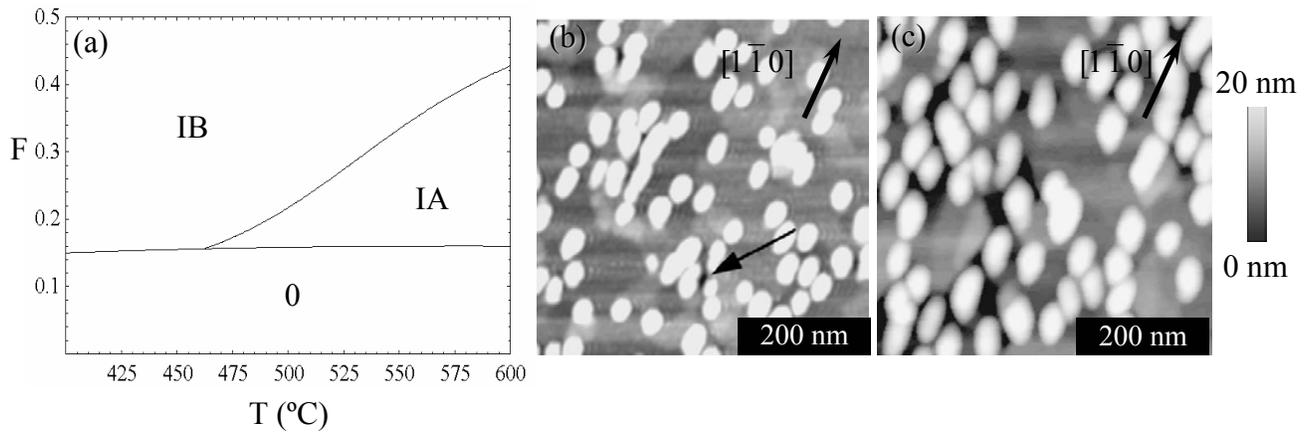

*Figure 8*
*Bouville et al.*